\documentclass[prl,reprint,superscriptaddress,aps]{revtex4-1}
\usepackage[]{graphicx}
\usepackage[]{amsmath}
\usepackage{braket}
\usepackage{pdfpages}

\begin{document}

\title{Mechanical spin control of nitrogen-vacancy centers in diamond}

\date{\today}
\author{E. R. MacQuarrie}
\affiliation{Cornell University, Ithaca, NY 14853}
\author{T. A. Gosavi}
\affiliation{Cornell University, Ithaca, NY 14853}
\author{N. R. Jungwirth}
\affiliation{Cornell University, Ithaca, NY 14853}
\author{S. A. Bhave}
\affiliation{Cornell University, Ithaca, NY 14853}
\author{G. D. Fuchs}
\email{gdf9@cornell.edu}
\affiliation{Cornell University, Ithaca, NY 14853}

\maketitle

\textbf{As spin-based quantum technology evolves, the ability to manipulate spin with non-magnetic fields is critical --- both for the development of hybrid quantum systems and for compatibility with conventional technology. Particularly useful examples are electric fields, optical fields, and mechanical lattice vibrations. The last of these represents direct spin-phonon coupling, which has garnered fundamental interest as a potential mediator of spin-spin interactions \cite{bennett, AlbrechtQuantPh2013}, but could also find applications in high-stability inertial sensing. In this Letter, we demonstrate direct coupling between phonons and nitrogen-vacancy (NV) center spins in diamond by inducing spin transitions with mechanically-driven harmonic strain. The ability to control NV spins mechanically can enhance NV-based quantum metrology, grant access to all transitions within the spin-1 quantum state of the NV center, and provide a platform to study spin-phonon interactions at the level of a few interacting spins.}

NV center spins are a promising solid-state platform for quantum information science \cite{BernienNature2013,ChildressScience2006} and precision metrology. They are sensitive magnetometers \cite{yacobyRobust}, electrometers \cite{doldeEfield}, and thermometers \cite{awschalomThermo, lukinThermo} with exceptional spatial resolution due to their atomic size \cite{MaminScience2013,StaudacherScience2013}. Significant progress in integrating NV centers with microelectromechanical systems (MEMS) \cite{DUrsoNJP2011,KolkowitzScience2012,ovartchaiyapongAPL2012,RablNatPhys2010} has paved the way for studies of spins coupled to mechanical resonators.  In previous work, NV centers have been coupled to cantilevers using either a magnetic field gradient or by tuning the frequency of a magnetic spin transition. Here we present resonantly-driven spin manipulation of NV centers, without a magnetic field, using gigahertz-frequency mechanical (stress) waves. This work demonstrates direct spin-phonon interactions at room temperature as means for quantum control of spin.

\begin{figure}[ht]
\begin{center}
\begin{tabular}{c}
\includegraphics[width=8cm]{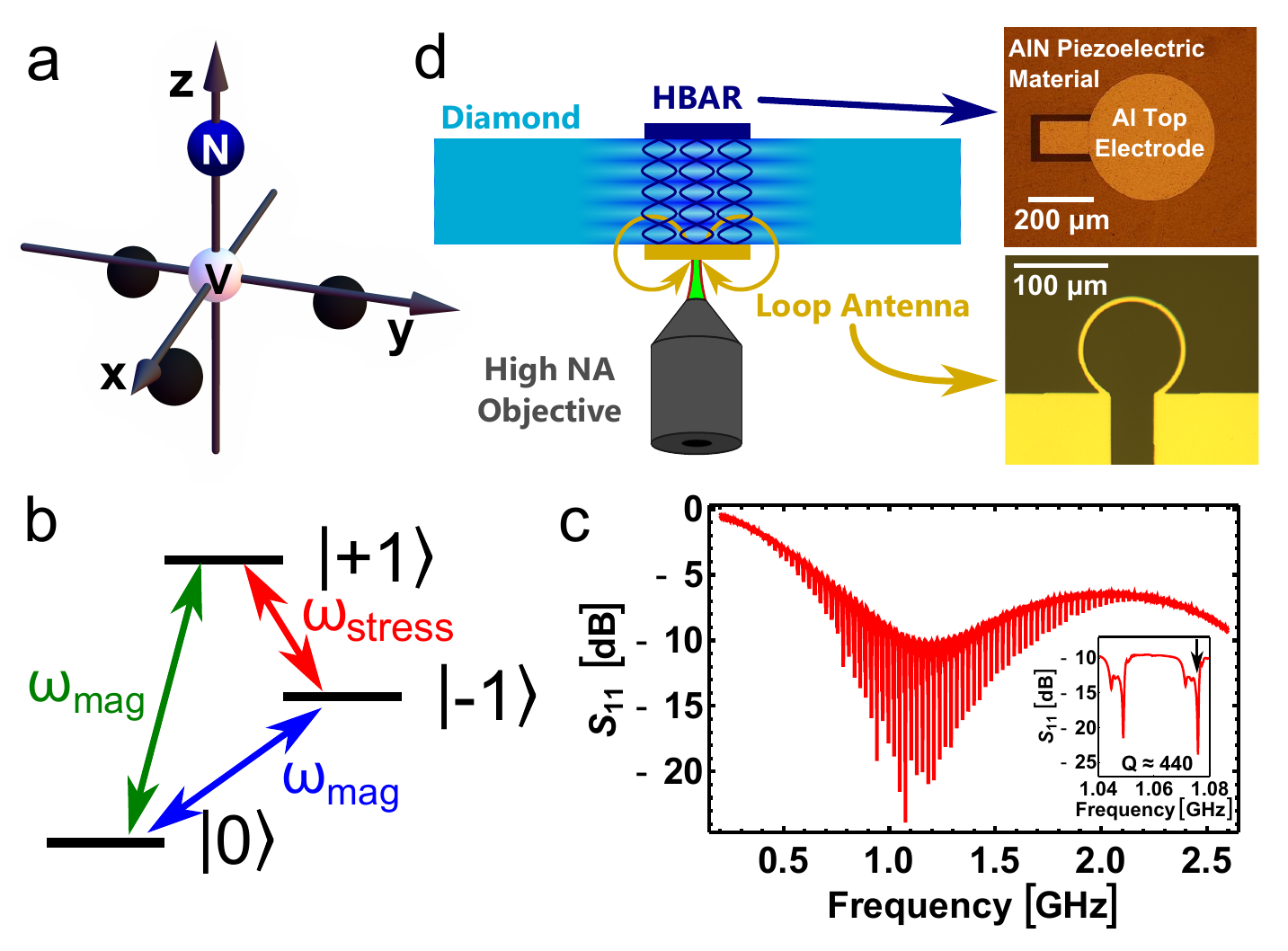} \\
\end{tabular} 
\end{center}
\caption[fig:fig1] {\textbf{Mechanical resonator on diamond for NV center spin control.} (a) Schematic of an NV center. The $z$-axis corresponds to the symmetry axis of the NV center; (b) Levels of an NV center ground-state spin. Magnetic driving enables $\Delta m_s = \pm 1$ transitions, whereas mechanical driving can produce $\Delta m_s = \pm 2$ transitions. (c) Reflected microwave power ($S_{11}$) as a function of frequency from the MEMS device measured using a network analyzer. Standing wave resonances have $Q$s as high as $437$; (d) Device schematic. A loop antenna produces gigahertz-frequency magnetic fields for magnetic control while a high-overtone bulk acoustic resonator (HBAR) produces gigahertz-frequency strain standing waves within the diamond.}
\label{fig:fig1}
\end{figure}

Driving spin transitions is the key to using NV center spins for quantum information science or sensing. Conventionally, quantum spin control in this system is accomplished with gigahertz-frequency magnetic fields \cite{fuchsReview,JelezkoPRL12004,Fuchs2009} or with optical fields at cryogenic temperature \cite{awschalomCPT}. Resonant lattice vibrations couple to nuclear quadrupole moments \cite{Bolef} and they represent another avenue to manipulate NV center electronic spins. NV centers couple to a magnetic field ($B_{\|}$ and $B_{\bot}$) and a perpendicular stress \textbf{$\sigma_{x/y}$} through their ground-state spin Hamiltonian \cite{doherty,doldeEfield}:
\begin{equation}
\begin{split}
H_{NV} &= D_{0}S_{z}^{2} + \gamma_{NV}B_{\|}S_{z}+ \gamma_{NV} B_{\bot} S_x \\
&+ \epsilon_{\bot}[\sigma_{x}(S_{x}S_{y}+S_{y}S_{x})+\sigma_{y}(S_{x}^{2}-S_{y}^{2})]
\end{split}
\label{eq:Hgs}
\end{equation}
where $D_{0}=2.87$~GHz is the zero-field splitting, $\gamma_{NV}=2.8$~MHz/G is the gyromagnetic ratio, $\epsilon_{\bot}=0.03$~MHz/MPa \cite{togan, bennett} is the perpendicular stress coupling constant, and ${S_x, S_y, S_z}$ are the $X$, $Y$, and $Z$ components of the spin-1 operator, respectively.  The $z$-axis is defined along the NV symmetry axis as depicted in Figure~\ref{fig:fig1}a.

In the $S_{z}$ basis, $H_{NV}$ has eigenstates $\left\{\Ket{(m_s=)0},\Ket{+1},\Ket{-1}\right\}$. $D_{0}$ breaks the degeneracy between the $\Ket{0}$ and $\Ket{\pm1}$ spin states at zero magnetic field. Careful alignment of the static external magnetic field $B_{\|}$ along the NV symmetry axis zeros the static component of $B_{\bot}$ and splits the $\Ket{+1}$ and $\Ket{-1}$ states. For conventional magnetic spin driving, an oscillating $B_{\bot}$ can drive spin transitions from the $\Ket{0}$ state to either the $\Ket{+1}$ or the $\Ket{-1}$ state. Similarly, a perpendicular stress couples the $\Ket{-1}$ and $\Ket{+1}$ states, allowing a direct \makebox{$\Ket{+1}\leftrightarrow\Ket{-1}$} spin transition to be driven by a gigahertz-frequency stress wave on resonance with the spin-state splitting. In the $S_z$ eigenbasis, this transition is magnetically forbidden by the magnetic dipole selection rule, $\Delta m_s=\pm1$. Thus, the ability to drive \makebox{$\Ket{+1}\leftrightarrow\Ket{-1}$} with an oscillating stress wave closes the loop on NV spin control in the ground state by establishing direct transitions between all three spin states, as depicted in Figure~\ref{fig:fig1}b.

The stress coupling coefficient $\epsilon_{\bot}$ is small enough that a large stress is required to produce a driving field comparable to those achieved with magnetic fields. To drive a large stress resonant with gigahertz-frequency spin transitions, we fabricated high-overtone bulk acoustic resonators~(HBARs) \cite{HBARref} on one face of a $\langle 100\rangle$ type IIa diamond that is dense with native NV centers ($\approx1.4\times10^{14}$~NV/cm$^{3}$). These MEMS transducers consist of a $3$~$\mu$m thick aluminum nitride (AlN) piezoelectric film sandwiched between two metal electrodes. Applying a gigahertz-frequency voltage across the AlN launches a longitudinal stress wave into the diamond. The diamond substrate acts as an acoustic Fabry-P\'{e}rot cavity, generating stress standing wave resonances with a pitch determined by the speed of sound in the diamond and its thickness. 

By measuring the microwave power reflected from the device ($S_{11}$), we observe the resonant frequency comb of an HBAR (Figure~\ref{fig:fig1}c). From this data, we used the Q-circle method \cite{qCircle} to find that the unloaded quality factors ($Q$s) of the resonances are as high as $Q=437$. Based on a one-dimensional oscillator model \cite{oneDosc}, this corresponds to a stress of \makebox{$\sigma_{max}\approx 10$~MPa} directed along the $\langle 100\rangle$ crystal axis of the diamond for 25 dBm applied microwave power. The component of the stress that is perpendicular to the NV symmetry axis is \makebox{$\sigma_{\bot}=\sqrt{2/3}\sigma_{max} \approx 8 $~MPa,} enough for a $\approx 240$~kHz spin driving field. On the opposite face of the diamond, we fabricated a loop antenna for magnetic spin control (Figure~\ref{fig:fig1}d). 

\begin{figure}[ht]
\begin{center}
\begin{tabular}{c}
\includegraphics[width=8cm]{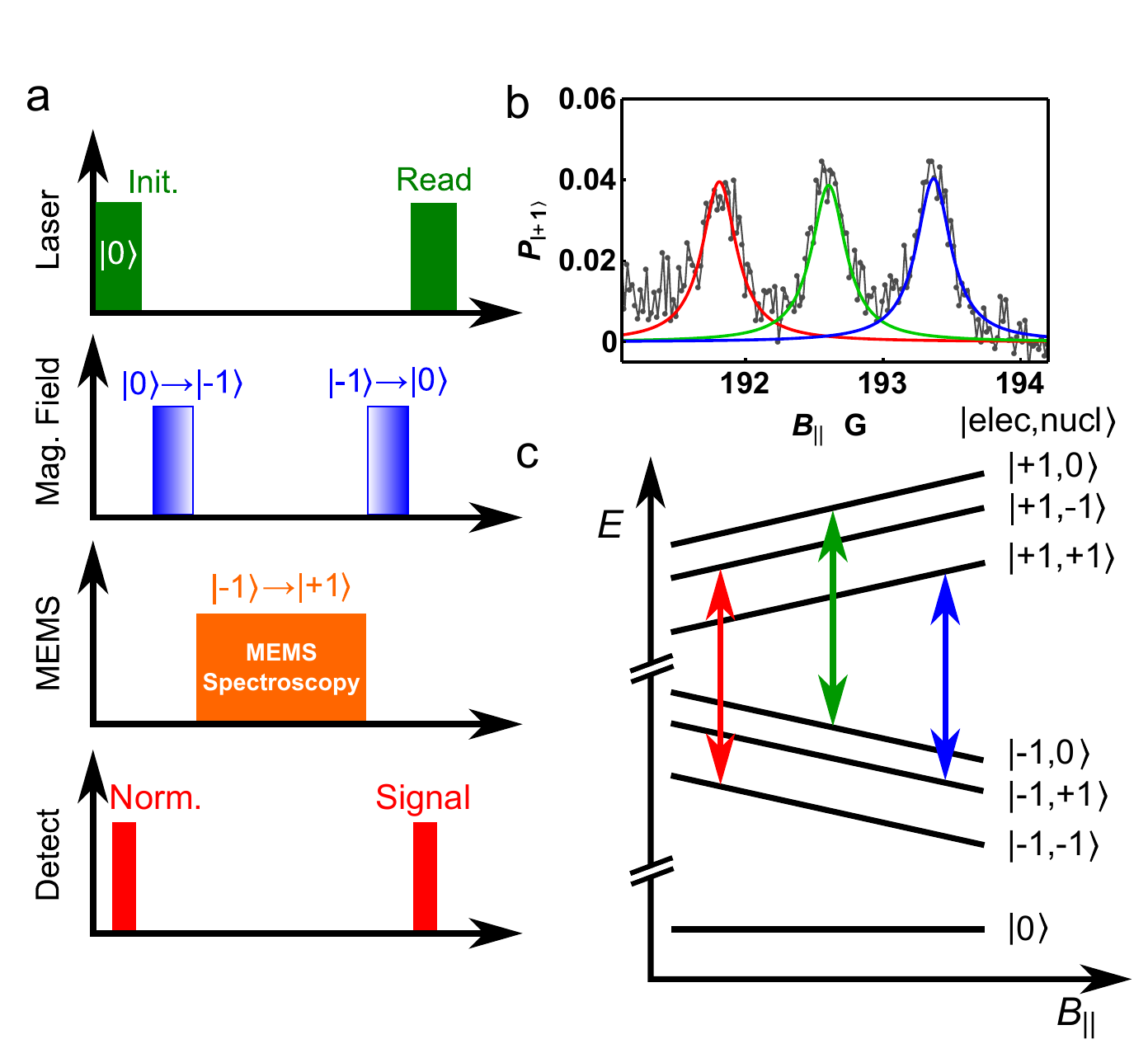} \\
\end{tabular}
\end{center} 
\caption[setup] {\textbf{Optically detected magnetic spin resonance.} (a) Pulse sequence used for ODMSR measurements; (b) Population driven into the $\Ket{+1}$ state by strain driving field as a function of the axial magnetic field $B_{\|}$ for \makebox{$\omega_{HBAR}=2 \pi \times 1.76~\text{GHz}$} at room temperature; (c) NV hyperfine structure labeled with the experimentally observed transitions.  Each arrow corresponds with the resonance condition \makebox{$\omega_{\pm 1}=\omega_{HBAR}$} for each of the three nuclear spin sublevels. }
\label{fig:fig2}
\end{figure}

To demonstrate mechanical spin control, we performed optically detected mechanical spin resonance (ODMSR) measurements of the $\Ket{-1}\rightarrow\Ket{+1}$ spin transition. The pulse sequences used for this experiment are shown in Figure~\ref{fig:fig2}a. First, the NV ensemble is initialized into $\Ket{0}$ by optical pumping with a $532$~nm laser. The laser is then turned off and a magnetic adiabatic passage through the $\Ket{0}\rightarrow\Ket{-1}$ resonance robustly transfers the initialized spin population into the $\Ket{-1}$ state \cite{Slicter1990}. The stress wave is then turned on for $6$~$\mu$s at a frequency $\omega_{HBAR}$ corresponding to a resonance of the HBAR. After this stress pulse, a second magnetic adiabatic passage transfers the population remaining in $\Ket{-1}$ to the $\Ket{0}$ state. Fluorescence read out of the population in the $\Ket{0}$ state is then performed, giving the signal for the experiment. Fluorescence read out is also performed after initialization into the $\Ket{0}$ state to provide normalization for each iteration of the duty cycle. By repeating this sequence as a function of $B_{\|}$, we scan $\omega_{\pm 1}$, the splitting energy between ${\Ket{+1}}$ and ${\Ket{-1}}$. Whenever \makebox{$\omega_{\pm 1}=\omega_{HBAR}$}, the strain pulse transfers population from $\Ket{-1}$ to $\Ket{+1}$. Population transferred to $\Ket{+1}$ during the stress pulse shows up as missing population in $\Ket{0}$ via fluorescence measurement at the end of the duty cycle. 

Typical ODMSR results are shown in Figure~\ref{fig:fig2}b. The spectrum shows three peaks with $0.78\pm0.02 $~G spacing. This corresponds to the \makebox{$A/\gamma_{NV}=0.77$~G} hyperfine splitting arising from interactions between the NV spin and the unpolarized nuclear spin of the $^{14}$N atom neighboring the vacancy \cite{doherty}. To account for dephasing and inhomogeneous driving, we calibrate the spin contrast by driving with conventional magnetic spin resonance \cite{SI}. For this resonance at $\omega_{HBAR}=1.076$~GHz, we estimate the peak strain driving field is $\approx230$~kHz. This is consistent with the coupling strength of $0.03$~MHz/MPa, which was previously determined from measurements of static strain at low temperature \cite{togan,bennett}. Measurements done at a different HBAR resonance shows that the stress driving field scales with Q \cite{SI}.

Because both stress and electric fields enter the NV spin Hamiltonian in the same way, we checked to see if the ODMSR signals result from stray electric fields. To address this possibility, we used the finite element analysis software ANSYS HFSS to simulate the electric field generated during the stress pulse, which comprises the dominant source of stray electric field in the experiment. The loop antenna on the rear face of the diamond was included. The simulation of the electric field within the relevant region of the diamond was no larger than $E_{sim}=10$~V/cm. The coupling strength between a perpendicular electric field and the NV ground state spin is $d_{gs}^{\bot}=17\pm3$~Hz cm/V \cite{vanoort}. Under conservative assumptions, $E_{sim}$ would generate a driving field roughly four orders of magnitude lower than observed in the experiment. We also considered, but ruled out, magnetic driving of the \makebox{$\Ket{+1}\leftrightarrow\Ket{-1}$} transition via stray magnetic fields from the MEMS transducer \cite{SI}.

%First order perturbation theory shows that nonaxial static magnetic fields couple the $\Ket{\pm1}$ and $\Ket{0}$ states \cite{SI} so the combination of a misaligned $B_{\|}$ and a resonant $B_{\bot}$ could also drive the \makebox{$\Ket{-1}\rightarrow\Ket{+1}$} spin transition. The coupling induced by such a misaligned $B_{\|}$ is weak, however, and $B_{\bot}\approx10???$~kHz is needed to produce the ODMSR signal we observe. Since our loop antennae at full power are only capable of producing fields on the order of $10$~MHz, it is unlikely that stray magnetic fields are driving our ODMSR signal. 

\begin{figure}[ht]
\begin{center}
\begin{tabular}{c}
\includegraphics[width=8cm]{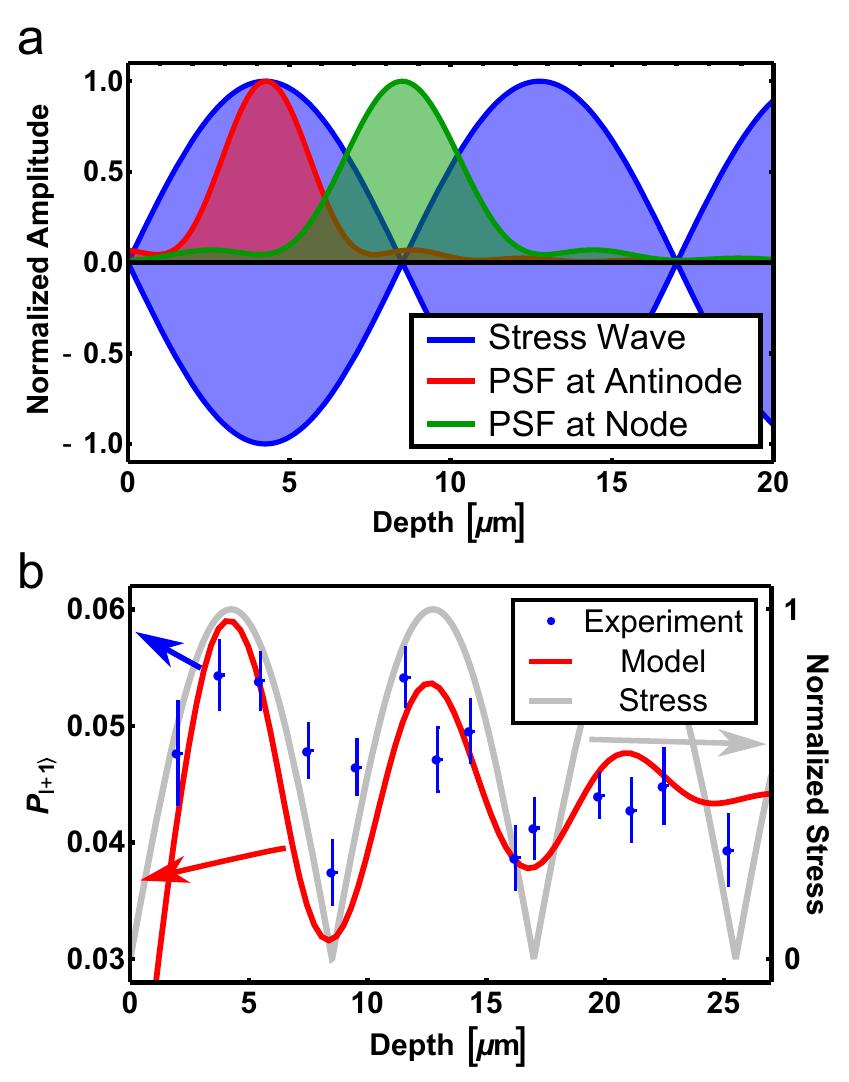} \\
\end{tabular}
\end{center} 
\caption[setup] {\textbf{Spatial periodicity of mechanical stress.} (a) Normalized point spread function (PSF) of our confocal microscope plotted at a node and an anti-node of the stress standing wave; (b) Peak ODMSR signal as a function of depth inside the diamond. The oscillations as a function of focal depth correspond to oscillations in stress along the standing wave used to drive spin transitions. }
\label{fig:fig3}
\end{figure}

As a critical verification that we drive spin transitions with mechanically-generated stress waves, we investigated how the ODMSR signal varied as a function of depth. Because we drive a stress standing wave, we expect that the ODMSR signal will be modulated at the periodicity of the standing wave. Taking care to account for aberrations introduced into our microscope from refraction at the air-diamond interface \cite{stelzer}, we repeated ODMSR measurements at different depths within the sample. To correctly interpret the results, we note that our microscope collects fluorescence from all of the NV centers within its confocal volume. Figure~\ref{fig:fig3}a depicts schematically the variation in stress amplitude across an approximate confocal point spread function (PSF) for our microscope. Different regions within the confocal volume experience different stress driving fields, and we sample the range of stress within the PSF. This reduces the spatial resolution in the focal direction and contributes to the inhomogeneous spin dephasing of the NV centers.

The ODMSR signal derives from the overlap between oscillating stress and the PSF of the microscope, which is a maximum at an anti-node of the stress standing wave.  In contrast, ODMSR measured at a node is reduced by a factor of $1.5$ (Figure~\ref{fig:fig3}b). Our approximate model, which is not a fit to the data, reproduces this ratio and the structure of the oscillation. To calculate the model curve, we convolve the PSF with the standing wave, while accounting for distortions in the position and shape of the PSF as a function of focal depth within the diamond. Crucially, we find excellent agreement between the spatial periodicity of the ODMSR signal and the wavelength of the HBAR-generated standing wave, calculated from the speed of sound in diamond, $1.6$~km/s, which is determined from the HBAR resonance pitch and the sample thickness. The decay of the ODMSR oscillation at large depth is due to spherical aberration of the optical PSF, which increases linearly as a function of depth. Taken together, these observations are the `smoking-gun' for mechanically-driven spin transitions because the stress standing wave is the only element of the experiment with spatial periodicity. 

The modest ODMSR amplitudes of these measurements are limited by the amplitude of the stress wave in this first generation of devices. Improvements in the fabrication is estimated to increase the HBAR $Q$s by greater than a factor of five at room temperature, and cooling the samples to cryogenic temperature can increase the $Q$ by a factor of $\approx 10^3$ \cite{TiffanyMEMS}. Either of these modest engineering improvements, along with improvements in the power handling capability, will put stress driving fields on an equal footing with magnetic driving and enable the use of stress driving for coherent control over single NV center spins and spin ensembles. 

Such control also has a number of practical sensing applications. Fang \textit{et al.} demonstrated that NV center magnetometry using a $\Ket{+1}$ and $\Ket{-1}$ qubit is more sensitive than a conventional $\Ket{0}$ and $\Ket{+1}$ or $\Ket{-1}$ qubit \cite{beausoleilBeats}. In that work, measurements were constrained to $B \sim 0$ so that magnetic driving pulses could magnetically decouple the nearly degenerate $\Ket{\pm 1}$ states through the $\Ket{0}$ state. Mechanical driving, however, enables direct decoupling of the $\Ket{\pm1}$ qubit at arbitrary $B_{\|}$. Also, spin-phonon driving of NV centers coupled to a mechanical resonator at low temperatures enables NV spin-state squeezing, which could push the sensitivity of NV magnetometry below the projection noise limit \cite{bennett}. NV center spins could also be used as a precision strain sensor integrated with a MEMS accelerometer, to compliment recent proposals to use NV center spins in gyroscopes \cite{cappellaro, budkerInertial}, with the aim of augmenting the long-term stability of MEMS-based inertial sensing.  

The development of new technology based on quantum spins in the solid-state will depend on integration with both existing technology and other qubit systems. We have demonstrated quantum manipulation through a direct interaction between spin and resonantly-driven cavity phonons, thus providing a new tool for integration and a new  avenue for fundamental studies of coherently-driven phonons interacting with a single spin. 

\section{Methods}

\subsection{Fabrication}

The substrate is an `optical-grade,' $300 \mu$m thick, single-crystal diamond purchased from Element Six. The sample was first cleaned in a boiling nitric:sulfuric (3:2) acid solution for one hour. Ti/Pt ($25$/$225$~nm) electrodes were then patterned on one face to serve as the ground plane for the HBARs. AlN was sputtered to a thickness of $3$~$\mu$m, and an Al layer was patterned on top to serve as the HBAR signal electrode. The HBAR was then coated in photoresist for protection while Ti/Pt ($25$/$225$~nm) loop antennas were patterned on the opposite face. 

\subsection{PSF Analysis}

Refraction at the air-diamond interface distorts the point spread function (PSF) of our confocal microscope and shifts the focal position deeper into the diamond. For the $0.8$~NA objective used here, this shift is described by the expression $d_{dia} \cong 3.1 d_{air}$ \cite{SI}. The distortion of the wavefront can be modeled by a FWHM that increases with depth \cite{SI}. Measurements of single NV centers at shallow depths place the FWHM of our microscope PSF at $\approx2$~$\mu$m near the diamond surface. Using this as a starting point, we modeled the variation in the PSF as a function of focal depth as a sinc function \cite{bornAndWolf} with a peak position and FWHM that shift as described above. By normalizing this PSF, convolving it with a standing wave, and normalizing the result to the integrated area of the PSF, we were able to produce the theoretical curve shown in Figure~\ref{fig:fig3}b. It is important to note that this is approximate because it is not calculated using full diffraction theory and does not include non-refractive aberrations.  Therefore, some differences between the model and the experiment are expected.

\vspace*{4mm}

% acknowledgement.tex
%
This work was supported by the Cornell Center for Materials Research with funding from the NSF MRSEC program (DMR-1120296) and the Department of Energy Office of Science Graduate Fellowship Program (DOE SCGF), made possible in part by  the American Recovery and Reinvestment Act of 2009, administered by ORISE-ORAU under contract no. DE-AC05-06OR23100.  Device fabrication was performed in part at the Cornell NanoScale Science and Technology Facility, a member of the National Nanotechnology Infrastructure Network, which is supported by the National Science Foundation (Grant ECCS-0335765).

\bibliography{bibDiMEMS}{}

\begin{thebibliography}{10}

\bibitem{bennett}
Bennett, S.~D., Yao, N.~Y., Otterbach, J., Zoller, P., Rabl, P., and Lukin,
  M.~D.
\newblock {\em Phys. Rev. Lett.}{ \bf 110}(156402) (2013).

\bibitem{AlbrechtQuantPh2013}
Albrecht, A., Retzke, A., Jelezko, F., and Plenio, M.~B.
\newblock {\em arXiv:}{ \bf }, 1304.2192 (2013).

\bibitem{BernienNature2013}
Bernien, H., Hensen, B., Pfaff, W., Koolstra, G., Blok, M.~S., Robledo, L.,
  Taminiau, T.~H., Markham, M., Twitchen, D.~J., Childress, L., and Hanson, R.
\newblock {\em Nature}{ \bf 497}, 86--90 (2013).

\bibitem{ChildressScience2006}
Childress, L., Gurudev~Dutt, M.~V., Taylor, J.~M., Zibrov, A.~S., Jelezko, F.,
  Wrachtrup, J., Hemmer, P.~R., and Lukin, M.~D.
\newblock {\em Science}{ \bf 314}(5797), 281--285 (2006).

\bibitem{yacobyRobust}
Maletinsky, P., Hong, S., Grinolds, M.~S., Hausmann, B., Lukin, M.~D.,
  Walsworth, R.~L., Loncar, M., and Yacoby, A.
\newblock {\em Nat. Nanotechnol.}{ \bf 7}, 320--324 (2012).

\bibitem{doldeEfield}
Dolde, F., Fedder, H., Doherty, M.~W., N\"{o}bauer, T., Rempp, F.,
  Balasubramanian, G., Wolf, T., Reinhard, F., Hollenberg, L. C.~L., Jelezko,
  F., and Wrachtrup, J.
\newblock {\em Nature Phys.}{ \bf 7}, 459--463 (2011).

\bibitem{awschalomThermo}
Toyli, D.~M., de~las Casas, C.~F., Christle, D.~J., Dobrovitski, V.~V., and
  Awschalom, D.~D.
\newblock {\em Proc. Natl. Acad. Sci. USA}{ \bf 110}(8417) (2013).

\bibitem{lukinThermo}
Kucsko, G., Maurer, P.~C., Yao, N.~Y., Kubo, M., Noh, H.~J., Lo, P.~K., Park,
  H., and Lukin, M.~D.
\newblock {\em arXiv:}{ \bf 1304.1068} (2013).

\bibitem{MaminScience2013}
Mamin, H.~J., Kim, M., Sherwood, M.~H., Rettner, C.~T., Ohno, K., Awschalom,
  D.~D., and Rugar, D.
\newblock {\em Science}{ \bf 339}(6119), 557--560 (2013).

\bibitem{StaudacherScience2013}
Staudacher, T., Shi, F., Pezzagna, S., Meijer, J., Du, J., Meriles, C.~A.,
  Reinhard, F., and Wrachtrup, J.
\newblock {\em Science}{ \bf 339}(6119), 561--563 (2013).

\bibitem{DUrsoNJP2011}
D'Urso, B., Dutt, M. V.~G., Dhingra, S., and Nusran, N.~M.
\newblock {\em New Journal of Physics}{ \bf 13}(4), 045002 (2011).

\bibitem{KolkowitzScience2012}
Kolkowitz, S., Bleszynski~Jayich, A.~C., Unterreithmeier, Q.~P., Bennett,
  S.~D., Rabl, P., Harris, J. G.~E., and Lukin, M.~D.
\newblock {\em Science}{ \bf 335}(6076), 1603--1606 (2012).

\bibitem{ovartchaiyapongAPL2012}
Ovartchaiyapong, P., Pascal, L. M.~A., Myers, B.~A., Lauria, P., and Jayich, A.
  C.~B.
\newblock {\em Applied Physics Letters}{ \bf 101}(16), 163505 (2012).

\bibitem{RablNatPhys2010}
Rabl, P., Kolkowitz, S.~J., Koppens, F. H.~L., Harris, J. G.~E., and Lukin, P.
  Z. M.~D.
\newblock {\em Nature Physics}{ \bf 6}, 602--608 (2010).

\bibitem{fuchsReview}
Dobrovitski, V.~V., Fuchs, G.~D., Falk, A.~L., Santori, C., and Awschalom,
  D.~D.
\newblock {\em Annu. Rev. Cond. Mat. Phys.}{ \bf 4}, 23--50 (2013).

\bibitem{JelezkoPRL12004}
Jelezko, F., Gaebel, T., Popa, I., Gruber, A., and Wrachtrup, J.
\newblock {\em Phys. Rev. Lett.}{ \bf 92}, 076401 Feb  (2004).

\bibitem{Fuchs2009}
Fuchs, G.~D., Dobrovitski, V.~V., Toyli, D.~M., Heremans, F.~J., and Awschalom,
  D.~D.
\newblock {\em Science}{ \bf 326}, 1520 (2009).

\bibitem{awschalomCPT}
Yale, C.~G., Buckley, B.~B., Christle, D.~J., Burkard, G., Heremans, F.~J.,
  Bassett, L.~C., and Awschalom, D.~D.
\newblock {\em Proc. Natl. Acad. Sci. USA}{ \bf 110}(7595) (2013).

\bibitem{Bolef}
Bolef, D.~I. and Sundfors, R.~K.
\newblock {\em Nuclear Acoustic Resonance}.
\newblock Acedemic Press,  (1993).

\bibitem{doherty}
Doherty, M.~W., Dolde, F., Fedder, H., Jelezko, F., Wrachtrup, J., Manson,
  N.~B., and Hollenberg, L. C.~L.
\newblock {\em Phys. Rev. B}{ \bf 85}(205203) (2012).

\bibitem{togan}
Togan, E., Chu, Y., Trifonov, A.~S., Jiang, L., Maze, J., Childress, L., Dutt,
  M. V.~G., rensen, A. S.~S., Hemmer, P.~R., Zibrov, A.~S., and Lukin, M.~D.
\newblock {\em Nature}{ \bf 466}, 730--734 (2010).

\bibitem{HBARref}
Lakin, K.~M., Kline, G.~R., and McCarron, K.~T.
\newblock {\em Microwave Symposium Digest, 1993., IEEE MTT-S International}{
  \bf }, 1517--1520 (1993).

\bibitem{qCircle}
Feld, D.~A., Parker, R., Ruby, R., Bradley, P., and Dong, S.
\newblock {\em Ultrasonics Symposium, 2008. IUS 2008. IEEE}{ \bf }, 431--436
  (2008).

\bibitem{oneDosc}
Graff, K.~F.
\newblock {\em Wave Motion in Elastic Solids}.
\newblock Dover Publications,  (1991).

\bibitem{Slicter1990}
Slicter, C.~P.
\newblock {\em Princples of Magnetic Resonance}.
\newblock Springer, 3rd edition,  (1990).

\bibitem{SI}
See Supplementary Information.

\bibitem{vanoort}
Vanoort, E. and Glasbeek, M.
\newblock {\em Chem. Phys. Lett.}{ \bf 168} (1990).

\bibitem{stelzer}
Hell, S., Reiner, G., Cremer, C., and Stelzer, E. H.~K.
\newblock {\em J. Microsc.}{ \bf 169}(3), 391--405 (1993).

\bibitem{TiffanyMEMS}
Cheng, T., Hsiao, J., Bhave, S.~A., and Duwel, A.
\newblock {\em in preparation}{ \bf } (2013).

\bibitem{beausoleilBeats}
Fang, K., Acosta, V.~M., Santori, C., Huang, Z., Itoh, K.~M., Watanabe, H.,
  Shikata, S., and Beausoleil, R.~G.
\newblock {\em Phys. Rev. Lett.}{ \bf 110}(130802) (2013).

\bibitem{cappellaro}
Ajoy, A. and Cappellaro, P.
\newblock {\em Phys. Rev. A}{ \bf 86}(062104) (2012).

\bibitem{budkerInertial}
Ledbetter, M.~P., Jensen, K., Fischer, R., Jarmola, A., and Budker, D.
\newblock {\em Phys. Rev. A}{ \bf 86}(052116) (2012).

\bibitem{bornAndWolf}
Born, M. and Wolf, E.
\newblock {\em Principles of Optics: Electromagnetic Theory of Propagation,
  Interference and Diffraction of Light}.
\newblock Cambridge University Press,  (1999).

\end{thebibliography}
\bibliographystyle{nature}

% start of the SI

\newpage

\begin{center}
\textbf{\large Mechanical spin control of nitrogen-vacancy centers in diamond: Supplementary Information}
\end{center}

%\date{\today}
%\author{E. R. MacQuarrie}
%\affiliation{Cornell University, Ithaca, NY 14853}
%\author{T. A. Gosavi}
%\affiliation{Cornell University, Ithaca, NY 14853}
%\author{N. R. Jungwirth}
%\affiliation{Cornell University, Ithaca, NY 14853}
%\author{S. A. Bhave}
%\affiliation{Cornell University, Ithaca, NY 14853}
%\author{G. D. Fuchs}
%\email{gdf9@cornell.edu}
%\affiliation{Cornell University, Ithaca, NY 14853}

\section{Experimental Details}

\begin{figure}[ht]
\begin{center}
\begin{tabular}{c}
\includegraphics[width=8cm]{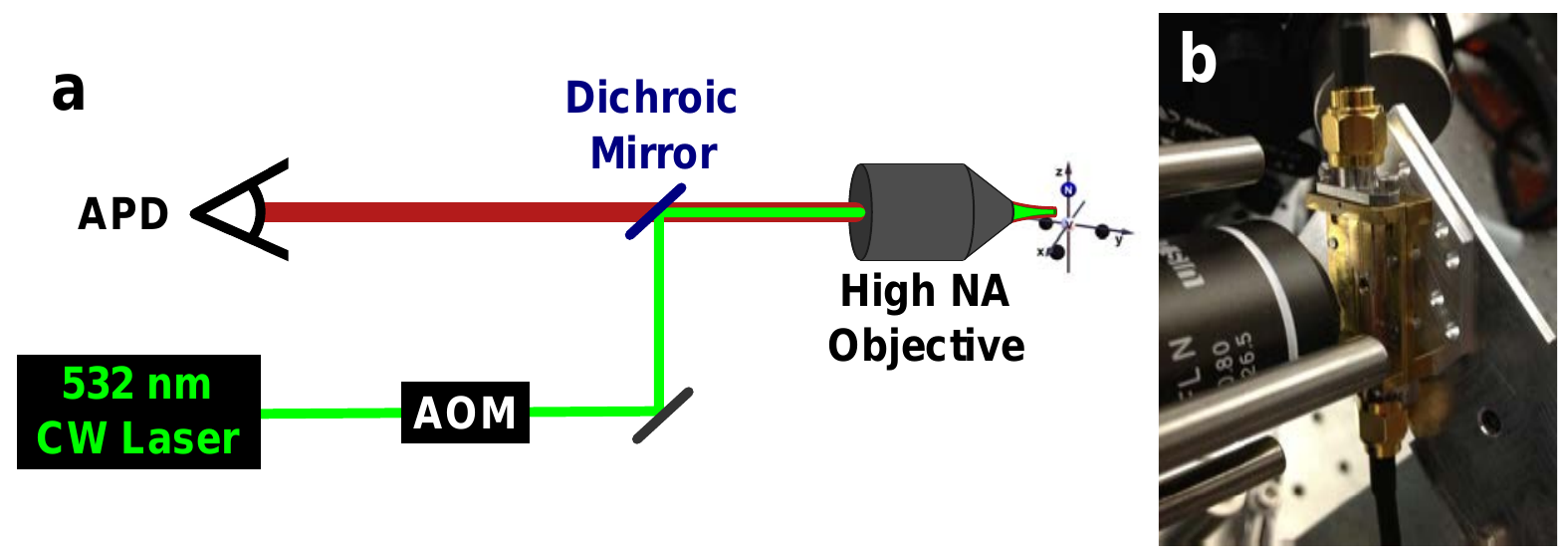} \\
\end{tabular}
\end{center} 
\caption[setup] {(a) Home-built confocal microscope; (b) Diamond sample mounted in the microscope with custom-built brass sample box.  The front side of the diamond (facing the microscope objective) has a lithographically-patterned microwave antenna wire bonded to a coplanar waveguide within the brass sample box. The back side contains the AlN transducer, which is also wire-bonded to a coplanar waveguide built into the brass sample box.  Behind the brass sample box is visible the permanent magnet, which sources the static magnetic field in the experiment.   }
\label{fig:figS1}
\end{figure}

Measurements were performed using a home-built confocal microscope (Figure~S\ref{fig:figS1}a). An Oxxius SLIM-532 $150$~mW laser was focused through a Gooch \& Housego 15210 acousto-optic modulator (AOM) that was used as a high-speed shutter. An Optics in Motion 101 fast-scanning mirror was used control the lateral position of the confocal focus. Both excitation and NV emission were focused through an Olympus LMPLFLN, 100x objective. The emission was detected with a Excelitas SPCM-AQRH-FC avalanche photodiode. 

Samples were mounted in the custom-built brass sample box pictured in Figure~S\ref{fig:figS1}b. The high-overtone bulk acoustic resonator (HBAR) device was powered by a Stanford Research Systems SG384 signal generator amplified by a Minicircuits ZRL-1150LN amplifier. The loop antenna was powered by a Tektronix 7122B Arbitrary Wave Generator (AWG) amplified by an Ophir 5161FE amplifier. A Stanford Research Systems DG645 digital delay generator triggered by the AWG was used to synchronize the various instruments and pulse the AOM. The axial magnetic field $B_{\|}$ was produced by a NdFeB permanent magnet on a motorized translation stage to enable field scanning.

\section{Driving Field Calibration}

We used conventional pulsed magnetic resonance signals to estimate the stress driving fields associated with the ODMSR signal. For calibration we modified the pulse sequence shown in Figure~2a of the main text by replacing the stress pulse with a magnetic field pulse resonant with the $\Ket{-1}\rightarrow\Ket{0}$ spin transition. The population that was driven into the $\Ket{0}$ state by this pulse was returned to the $\Ket{-1}$ state during the second adiabatic passage, allowing us to observe spin driving as an absence of population in the $\Ket{0}$ state. 

We tuned the amplitude of the magnetic field pulse until the amplitude of the magnetic resonance signal matched that of our ODMSR signals. At this point, the driving fields generated by the stress pulse and the magnetic field pulse are equal, giving us a point of comparison. Because the magnetic driving field scales as $\sqrt{P}$ where $P$ is the microwave power coupled into the loop antenna, we measured the magnetic driving field directly at a higher value of $P$ by observing Rabi oscillations, and extrapolated downward to find the driving field generated by the weaker magnetic pulse. 

To estimate the total driving amplitude of the ODMSR signals, we first subtract a constant background from the ODSMR signal that we attribute to pulse errors. We then added the amplitudes of the three hyperfine peaks displayed in Figure~2b of the main text. The inhomogeneities in the strain driving field are modeled with the overlap of the laser PSF and the stress standing wave. For the $1.076$~GHz stress resonance at an anti-node, this is a factor of $0.84$. Dividing the sum of the peak amplitudes by this correction factor gives the ideal ODMSR signal that we compare with our magnetic resonance data. In this way, we determine the peak stress driving field to be 230~kHz. 

\section{Measurements at Different $Q$ Values}

Measurements were taken at two different stress resonances ($\omega_{1}=1.076$~GHz and $\omega_{2}=1.103$~GHz)which had $Q$'s of $437$ and $350$ respectively. Since we are not driving the $\Ket{-1}\rightarrow\Ket{+1}$ spin transition to saturation, the stress driving field is expected to depend linearly on the $Q$ of the resonance. To compare the driving field generated at different HBAR resonances, we take into account both the quality factor ($Q$) and impedance of the HBAR for each resonance.  The ratio of driving fields should match the ratio of stress, which may be calculated as 
\begin{equation}
r_{stress}=\frac{Q_1}{Q_2}\times\frac{V_1}{V_2}=\frac{Q_1}{Q_2}\times\frac{Z_1 \sqrt{50~\Omega+Z_2}}{Z_2 \sqrt{50~\Omega+Z_1}}
\label{eqn:r}
\end{equation}
where $Z_1=29.9~\Omega$ is microwave impedance of the HBAR at resonance 1 and $Z_2=33.5~\Omega$ is microwave impedance of the HBAR at resonance 2.  Using the two resonances mentioned above, we measure the ratio of driving fields to be $1.10 \pm 0.05$, which is close to the expected value of $r_{stress}=1.14$, which we calculate from equation~S\ref{eqn:r} and network analyzer measurements. 

\section{Rabi Driving with Strain}

\begin{figure}[ht]
%\begin{center}
%\begin{tabular}{c}
\includegraphics[width=8cm]{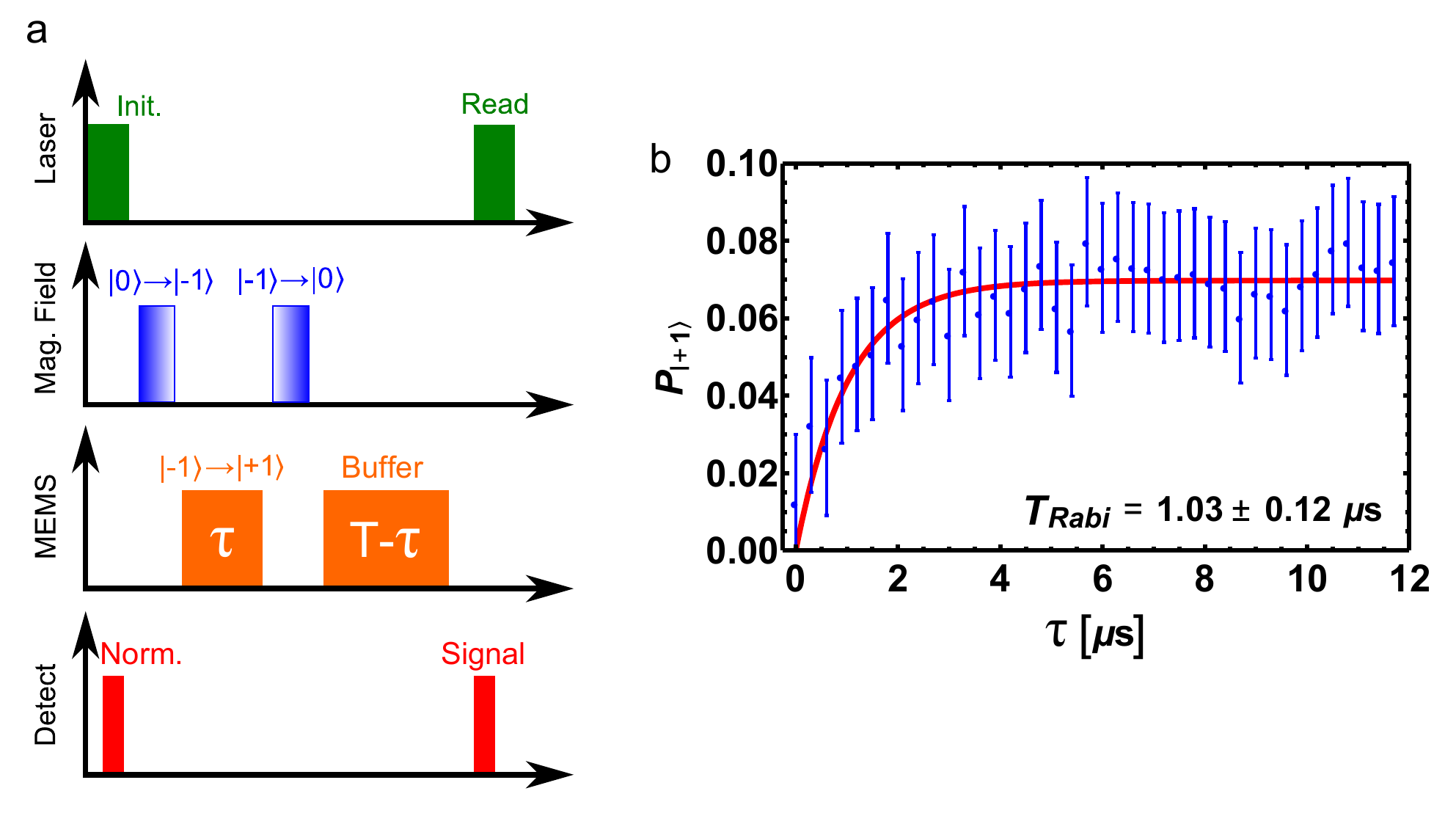}% \\
%\end{tabular}
%\end{center} 
\caption[setup] {(a) Pulse sequence for Rabi-style ODSMR measurement.  The first MEMS pulse is for state transition, whereas the second pulse keeps the total MEMS power fixed throughout the duty cycle.  The second pulse does not disturb the population of $\Ket{0}$, which is proportional to the fluorescence contrast in spin measurement.  (b) Results from Rabi-style ODSMR measurement.  No coherent oscillations are observed because we measure fluorescence from inhomogeneous strains within the confocal volume of our microscope.}
\label{fig:rabi}
\end{figure}

In order to determine the optimal pulse length for the stress wave in our measurements, we performed a Rabi-style ODSMR measurement where the length of the stress pulse $\tau$ was varied from $0$ to a maximum value $T$. The pulse sequence for this experiment is depicted in Figure~S\ref{fig:rabi}a. To mitigate thermal effects, the total power to the sample was kept constant by applying a second stress pulse with length $(T-\tau)$ after the second magnetic adiabatic passage (AP) has driven the $\Ket{-1}$ population back into the $\Ket{0}$ spin state. Because there is no magnetic pulse following this second stress pulse, population driven between $\Ket{+1}$ to $\Ket{-1}$ during the second pulse has no effect on fluorescence measurement, which is sensitive only to the population of $\Ket{0}$. 
The experimental signal, shown in Figure~S\ref{fig:rabi}b, decays exponentially in $\tau$.  The characteristic decay time is $T_{Rabi}=1.03\pm0.12$~$\mu$s, which we attribute to dephasing of the spin ensemble by inhomogeneous stress within the collection volume. 

\section{The effect of stray magnetic fields from the HBAR}

We examined the effect of stray \emph{magnetic} fields produced by the HBAR and resonant with the $\Ket{-1}$ $\Ket{+1}$ transition, as a potential spurious contribution to the ODMSR signal.  To understand this effect, we consider the Hamiltonian for an NV center in a static magnetic field that is aligned with the NV symmetry axis
\begin{equation}
H_{0}=D_{0}S_{z}^{2}+\gamma_{NV}B_{\|}S_{z},
\end{equation}
with the perturbation 
\begin{equation}
\Delta H=\gamma_{NV}B_{x0}S_{x},
\end{equation}
which represents a small ($B_{x0}\ll B_{\|}$) perpendicular component to the static magnetic field, aligned along the $x$-axis for simplicity. 

To first order, $\Delta H$ mixes the eigenstates of $H_{0}$ without shifting their energy. Rewriting $S_{x}$ in the basis defined by the first-order perturbed eigenstates, we find
\begin{widetext}
\begin{equation}
S_{x}'=\begin{pmatrix}
\frac{\gamma_{NV}B_{x0}}{\omega_{+1}} & \frac{1}{\sqrt{2}} & \frac{\gamma_{NV}B_{x0}}{2} \left(\frac{1}{\omega_{+1}} + \frac{1}{\omega_{-1}}\right) \\
\frac{1}{\sqrt{2}} & - \gamma_{NV}B_{x0} \left(\frac{1}{\omega_{+1}} + \frac{1}{\omega_{-1}}\right) & \frac{1}{\sqrt{2}} \\
 \frac{\gamma_{NV}B_{x0}}{2} \left(\frac{1}{\omega_{+1}} + \frac{1}{\omega_{-1}}\right) & \frac{1}{\sqrt{2}} & \frac{\gamma_{NV}B_{x0}}{\omega_{-1}}
\end{pmatrix},
\end{equation}
\end{widetext}
where we use $\omega_{+1}$ for the unperturbed $\Ket{0} \rightarrow \Ket{+1}$ transition energy and $\omega_{-1}$ for the unperturbed $\Ket{0} \rightarrow \Ket{-1}$ transition energy.  This can be re-written as 
\begin{equation}
\begin{split}
S_x'=S_x+\frac{\gamma_{NV} B_{x0}}{2} \left(\frac{1}{\omega_{+1}}+ \frac{1}{\omega_{-1}} \right)(S_x^2-S_y^2) \\ + \text{diagonal terms}. 
\end{split}
\label{eqn:SXp}
\end{equation} 
The second term in S\ref{eqn:SXp} introduces transverse anisotropy with the same form as perpendicular stress. Therefore, application of a resonant oscillating field, $B_{x1}(t)=B_{x1}\cos[(\omega_{+1}-\omega_{-1})t]$ weakly drives spin transitions in a mis-aligned static magnetic field. Under the rotating wave approximation, the $\Ket{-1}\leftrightarrow\Ket{+1}$ driving field will be
\begin{equation}
\begin{split}
\Omega_{\Ket{-1}\rightarrow\Ket{+1}}=\frac{\gamma_{NV}^{2}B_{x1}B_{x0}}{4} \left(\frac{1}{\omega_{+1}}+ \frac{1}{\omega_{-1}} \right)  \\  = \frac{\gamma_{NV}^{2}B_{x1}B_{x0} D_0}{2 (D_0^2-\gamma_{NV}^2 B_{\|}^2 )}.
\end{split}
\end{equation}
Comparing this with the conventional expression for driving the $\Ket{0}\rightarrow\Ket{\pm1}$ transition on resonance with ($\Omega_{\Ket{0}\rightarrow\Ket{\pm1}}=\gamma_{NV}B_{x1}/\sqrt{2}$), we find the following expression for the ratio of driving fields:
\begin{equation}
\frac{\Omega_{\Ket{-1}\rightarrow\Ket{+1}}}{\Omega_{\Ket{0}\rightarrow\Ket{\pm1}}}=\frac{\sqrt{2} \gamma_{NV}B_{x0}D_{0}}{D_{0}^{2}-\gamma_{NV}^{2}B_{\|}^2 }.
\label{eqn:magdrive}
\end{equation}
At $B_{\|}=192$~G, this results in a driving field of \makebox{$\Omega_{\Ket{-1}\rightarrow\Ket{+1}} \cong 0.0014 \times B_{x0} \times \Omega_{\Ket{0}\rightarrow\Ket{\pm1}}$}, where $B_{x0}$ has units of G.

\begin{figure}[ht]
\begin{center}
\begin{tabular}{c}
\includegraphics[width=8cm]{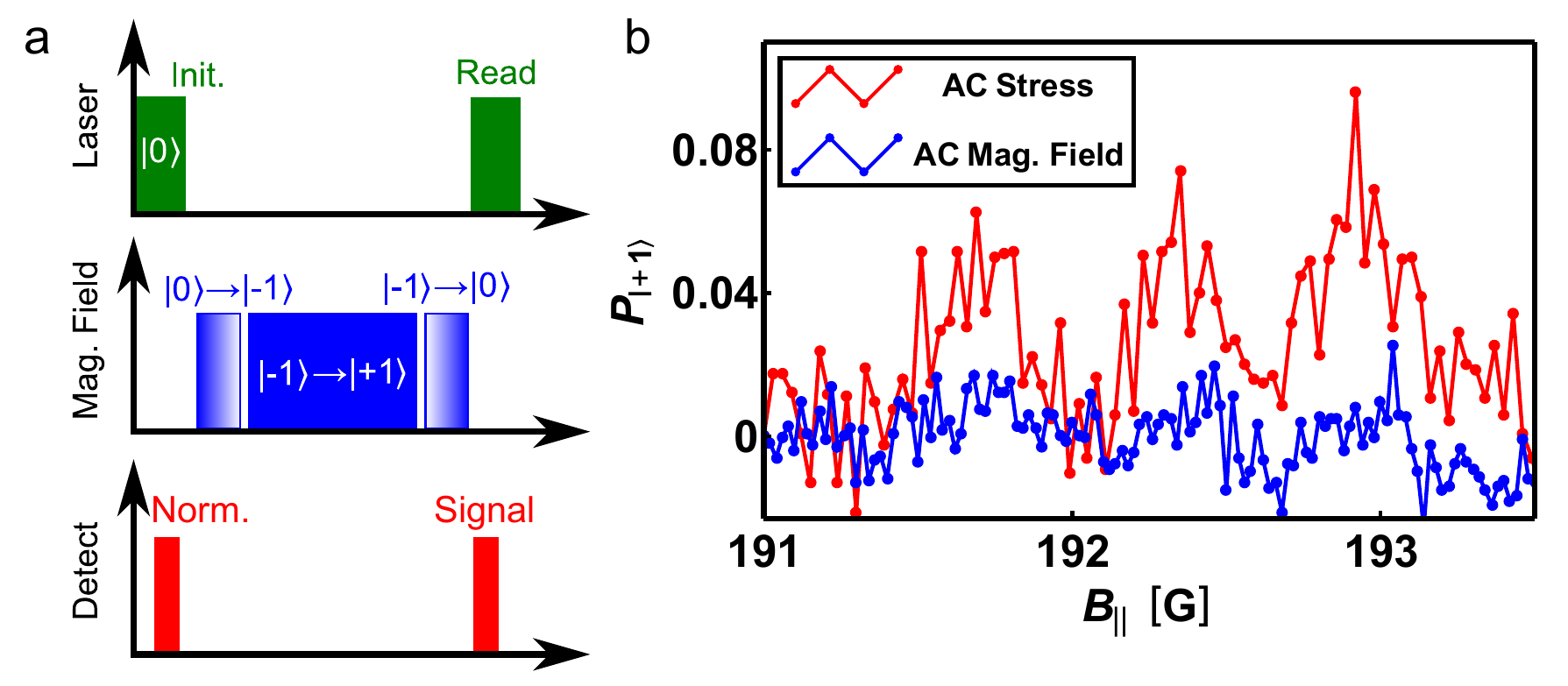} \\
\end{tabular}
\end{center} 
\caption[setup] {(a) Pulse sequence for measuring $B_{x1}$ field driving the $\Ket{-1}\rightarrow\Ket{+1}$ transition; (b) Magnetic driving of the $\Ket{-1}\rightarrow\Ket{+1}$ transition using $\Omega_{\Ket{0}\rightarrow\Ket{\pm1}} \approx 2 \pi \times 1 \text{MHz}$ (blue) plotted next to stress driving (red).  Both are measured on the 1.076 GHz acoustic resonance.}
\label{fig:control}
\end{figure}

As an experimental test of this effect, we replaced the stress wave pulse in our ODMSR experiments with a magnetic field pulse through the microwave antenna, resonant with the $\Ket{-1}\rightarrow\Ket{+1}$ transition, and with a power equivalent to $\Omega_{\Ket{0}\rightarrow\Ket{\pm 1}} = 2 \pi \times 1 \text{MHz}$ for conventional magnetic driving.  Because we measure in a slightly misaligned magnetic field ($B_{x0} \neq 0)$, we observed weak magnetic driving of the $\Ket{-1}\rightarrow\Ket{+1}$ transition. (Figure~S\ref{fig:control}b).

Using the electromagnetic simulation mentioned in the main text, we calculate that the stray magnetic field produced by the HBAR is $B_{1,HBAR}=0.17$~G at the depth of optical measurements, and directed perpendicular to the plane.  Under the most conservative estimate, we assume that $B_{x0}=10$~G.  Using equation S\ref{eqn:magdrive}, we estimate the stray magnetic field produces $\Omega_{\Ket{-1}\rightarrow\Ket{+1}}\cong 2 \pi \times 2.7$~kHz, about $100\times$ smaller than the stress-wave induced driving field.  This value is about $5\times$ smaller than our experimental test, where we intentionally introduced a driving field through the microwave antenna.  Interestingly, although we have just shown it is possible to drive the $\Ket{-1}\leftrightarrow\Ket{+1}$ spin transition magnetically, in most cases it is not practical. It requires very large values of $B_{x1}$ unless $B_{x0}$ is sizable. Unfortunately, as $B_{x0}$ grows, the eigenstates of the $S_{z}$ basis mix more strongly, reducing both NV read-out contrast and spin coherence [S1]. 

Magnetic driving of the $\Ket{-1}\leftrightarrow\Ket{+1}$ spin transition is fundamentally limited by weak coupling in the $S_z$ basis, but mechanical driving is not, where the primary limitation is the stress wave amplitude. Mechanical driving affects neither spin coherence nor read-out contrast. With room temperature HBAR  $Q$'s expected to improve by more than a factor of five in the next generation of devices, mechanical driving is the more practical for route quantum control using the $\Ket{-1}\leftrightarrow\Ket{+1}$ spin transition. 

\section{Corrections to PSF}

\begin{figure}[ht]
\begin{center}
\begin{tabular}{c}
\includegraphics[width=9cm]{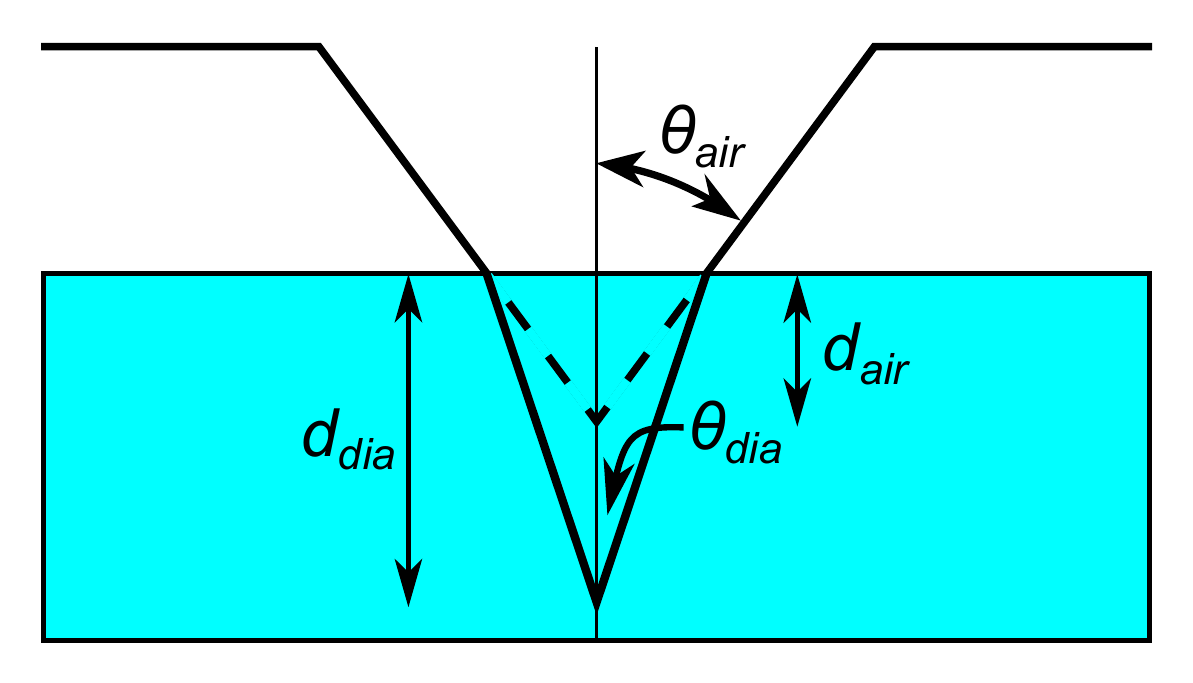} \\
\end{tabular}
\end{center} 
\caption[setup] {(a) Geometry used to calculate the aberration introduced by refraction at the air-diamond interface.}
\label{fig:focus}
\end{figure}

To accurately interpret our measurement of the stress standing wave (Figure~3b of main text), it was critical to know the shape and location of our microscope's point spread function (PSF) inside the diamond. This problem is nontrivial because refraction at the air-diamond interface introduces aberrations that shift the focus deeper into the diamond and increase the width of the focal spot (Figure~S\ref{fig:focus}). Geometric optics relates the depth in the diamond $d_{dia}$ to the nominal depth $d_{air}$ as
\begin{equation}
d_{dia}=\frac{n_{dia}\cos\theta_{dia}}{n_{air}\cos\theta_{air}} d_{air}
\label{eq:corr}
\end{equation}
where $n$ is the index of refraction and $n_{dia}\sin\theta_{dia}=n_{air}\sin\theta_{air}$. 

Assuming a constant intensity profile $I_{0}$, the power $P(r)$ that leaves the back of the objective is given by $P=\pi r^{2} I_{0}$. Expressing $P(r)$ as a function of the angle $\theta_{dia}$ and differentiating, we can use Equation~S\ref{eq:corr} to arrive at an expression for the differential power as a function of $\theta_{d}$
\begin{equation}
\frac{\textrm{d}P}{\textrm{d}\theta_{dia}}=2\pi I_{0} n_{dia}^{2}d_{air}^{2}\tan\theta_{dia}\sec^{2}\theta_{dia}\frac{\cos^{2}\theta_{dia}}{\cos^{2}\theta_{air}}.
\label{eq:dpdt}
\end{equation}
We can now use Equation~S\ref{eq:dpdt} as a weight to determine the average value of $d_{dia}$ as a function of $d_{air}$:
\begin{equation}
\left\langle d_{dia}\right\rangle= d_{air} \frac{\int^{\theta_{dia}^{max}}_{0} \frac{n_{dia}\cos\theta_{dia}}{n_{air}\cos\theta_{air}} \frac{\textrm{d}P}{\textrm{d}\theta_{dia}}}{\int^{\theta_{dia}^{max}}_{0} \frac{n_{dia}\cos\theta_{dia}}{n_{air}\cos\theta_{air}}} \approx 3.1d_{air}
\end{equation}
where $\theta_{dia}^{max}$ is defined by the expression 
\begin{equation}
\sin\theta_{dia}^{max}=\frac{\textrm{NA}}{n_{dia}}
\end{equation}
for $\textrm{NA}=0.8$.

This correction factor of $3.1$ serves as a lower bound to the true factor since we have not accounted for other aberrations in our microscope. For the wavelength of our measured standing wave to match the expected value of $17$~$\mu$m, the correction factor needs to be $3.4\pm0.1$. We attribute other aberrations, not accounted for in this calculation, for the small mismatch in the wavelength. The calculated $3.1$ correction factor was used at all times in the main text. 

In order to account for distortions to the shape of the PSF, we approximated the change in the PSF FWHM by the expression
\begin{equation}
\begin{split}
FWHM_{dia}=\Delta d_{dia}= \\ \frac{1}{2}\left(\frac{n_{dia}\cos\theta_{dia}^{max}}{n_{air}\cos\theta_{air}^{max}}-\frac{n_{dia}\cos\theta_{dia}^{min}}{n_{air}\cos\theta_{air}^{min}}\right)d_{air}
\end{split}
\end{equation}
where $\theta^{max}$ is set by the microscope NA and $\theta^{min}=0$. 

\noindent{[S1] Stanwix, P. L., Pham, L. M., Maze, J. R., Sage, D. L., Yeung, T. K., Cappellaro, P., Hemmer, P. R., Yacoby, A., Lukin, M. D., and Walsworth, R. L. Phys. Rev. B 82 (201201(R)) (2010).}

\end{document}